\documentclass{artificial-life}

\usepackage[style=apa,natbib=true]{biblatex}
\usepackage[inkscapelatex=false]{svg}
\usepackage{amsfonts}
\usepackage{amsmath}
\usepackage{subfig}
\usepackage{floatrow}
\usepackage{graphbox}
\usepackage{float}
\usepackage{xcolor}

\newcommand{\La}{\mathcal{L}}
\newcommand{\R}{\mathbb{R}}

\newcommand{\supp}{\mathcal{L}}
\newcommand{\website}{https://sites.google.com/view/flow-lenia}
\newcommand{\video}{https://sites.google.com/view/flowlenia/model}

\newcommand{\notebook}{https://colab.research.google.com/drive/1l-Og8xRlc5ew0489swuud0Me7Sc5bCss?usp=sharing}

\title{Flow-Lenia: Emergent evolutionary dynamics in mass conservative continuous cellular automata \footnote{This manuscript has been accepted for publication in the Artificial Life journal (\url{https://direct.mit.edu/artl}) \citep{plantec_flow-lenia_2025}}}

\auth{
    Erwan Plantec \affil{1}, % Note the escape for an underscore
    Gautier Hamon \affil{2},
    Mayalen Etcheverry \affil{2},
    Bert Wang-Chak Chan \affil{3},
    Pierre-Yves Oudeyer \affil{2},
    Clément Moulin-Frier \affil{2}
}

\corresponding{Erwan Plantec}{erpl@itu.dk}

\affiliations{
    \item IT University, REAL lab, Copenhagen, Denmark
    \item Flowers AI \& CogSci Lab, Inria Center of the University of Bordeaux, France 
    \item Google DeepMind, Tokyo, Japan
}

\abs{
Central to the artificial life endeavour is the creation of artificial systems spontaneously generating properties found in the living world such as autopoiesis, self-replication, evolution and open-endedness. While numerous models and paradigms have been proposed, cellular automata (CA) have taken a very important place in the field notably as they enable the study of phenomenons like self-reproduction and autopoiesis. Continuous CA like Lenia have been showed to produce life-like patterns reminiscent, on an aesthetic and ontological point of view, of biological organisms we call creatures. We propose in this paper \emph{Flow-Lenia}, a mass conservative extension of Lenia. We present experiments demonstrating its effectiveness in generating spatially-localized patters (SLPs) with complex behaviors and show that the update rule parameters can be optimized to generate complex creatures showing behaviors of interest. Furthermore, we show that Flow-Lenia allows us to embed the parameters of the model, defining the properties of the emerging patterns, within its own dynamics thus allowing for multispecies simulations. By using the evolutionary activity framework as well as other metrics, we shed light on the emergent evolutionary dynamics taking place in this system.}


\begin{document}

% Add animation of flow lenia in new website and change the link in section flow lenia system

% Don't Touch ! %%%%%% 
\coverpage           % 
%\linenumbers         % 
%\doublespacing       % 
%%%%%%%%%%%%%%%%%%%%%% 

%%% ALL TEXT OF ARTICLE BELOW THIS LINE %%%
%-----------------------------------------%

\begin{figure*}[h]
    \centering
  \includegraphics[width=.9\textwidth]{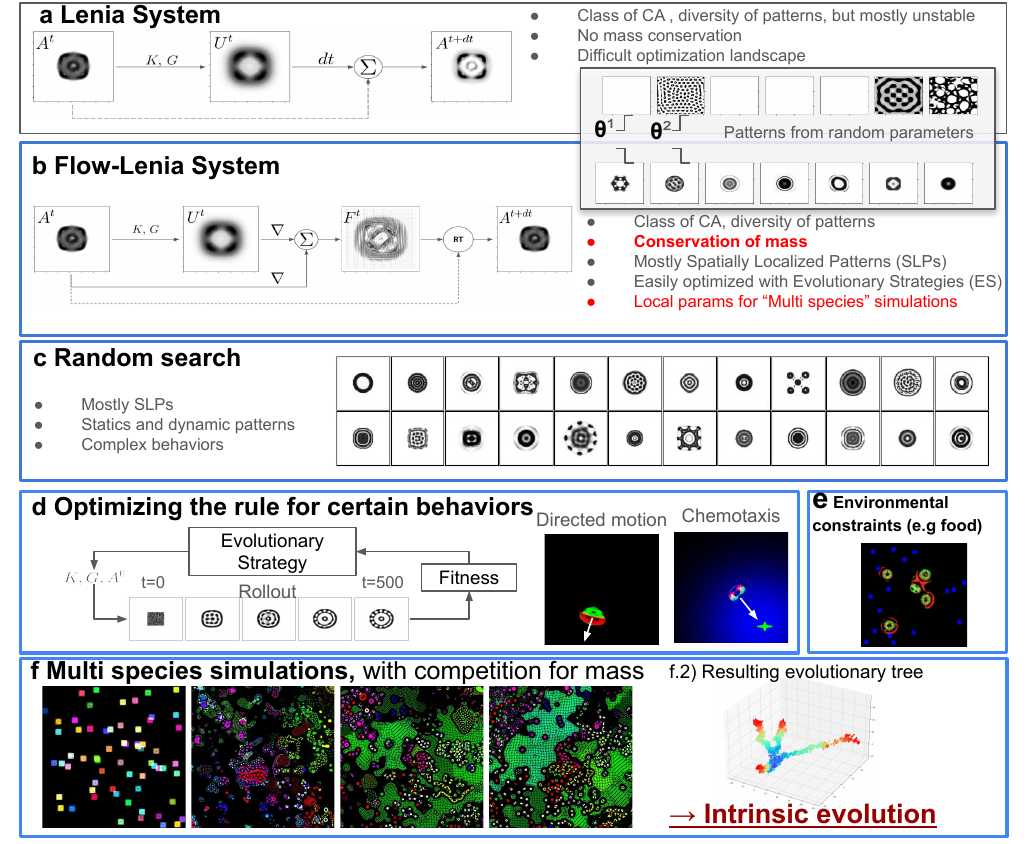}
    \caption{We present \emph{Flow-Lenia}, an extension of the Lenia (a) continuous Cellular Automata (CA). Flow-Lenia (b) introduces a built-in constraint for mass conservation, strongly facilitating the discovery of life-like patterns (c), the optimization of the system parameters towards certain behaviors (d) and the introduction of environmental constraints (e). Moreover, it allows to embed the system parameters within its own local dynamics, leading to large-scale multi-species simulations analysed in the light of the Evolutionary Activity framework (f). \textbf{(a) Lenia system}. The growth $U^t$ is computed with kernels $K$ and growth functions $G$. A small portion of the growth is then added to activations $A^t$ to give the next state $A^{t+dt}$. (section \ref{sec:Lenia}). \textbf{(b) Flow-Lenia system}. $U^t$ is computed as in Lenia and interpreted as an affinity map. The flow $F^t$ is given by combining the affinity map and activation gradients. The next state is obtained by ``moving'' matter in the CA space according to the flow $F^t$ using reintegration tracking. (section \ref{sec:FL}). Inset on the right of (a,b) shows 7 patterns obtained with randomly sampled update rules parameters, in Lenia (top, resulting mostly in non-SLP or empty patterns) and Flow-Lenia (bottom, resulting mostly in SLP patterns). \textbf{(c) Random search}. Patterns emerging from random parameter sampling in Flow-Lenia are qualitatively analyzed (sections \ref{meth:rs} and \ref{sec:RS}). \textbf{(d) Optimizing the system update rule} is performed using simple evolutionary strategies with respect to predefined fitness functions, resulting in creatures with specific behaviors (e.g directed motion or chemotaxis). (sections \ref{meth:ds} and \ref{sec:DS}). \textbf{(e) Environment constraints}. Example of environment with food (blue) that creatures can consume to gain mass. \textbf{(f) Multi species simulations}. Snapshots of a large scale multi-species simulation enabled by the parameter embedding mechanism (section \ref{sec:FLP}), resulting in a evolutionary tree (f.2) (section \ref{sec:Evo}).}
\label{fig:lenia-flowlenia}
\end{figure*}

\clearpage

\section{Introduction}

Cellular automata (CA), notably Conway's Game of Life (GoL) \citep{adamatzky2010}, have attracted a lot of interest from the artificial life (ALife) community because of the emergence of life-reminiscent spatially-localized patterns (SLPs). These patterns are of special interest as instances of autopoietic structures (i.e self-produced and self-maintained structures) \citep{beer2004}, a fundamental property of life and cognition as proposed in \citeauthor{maturana1991} theory \citep{maturana1991}. Recently, Lenia, a generalization of GoL to the continuous domain has been proposed \citep{chan2019,chan2020}. Previous studies with this system have shown the emergence of autopoietic SLPs, also called ``creatures'', often resembling microscopic life-forms and displaying various behaviors like motility or self-replication. Moreover, Lenia displays a much wider diversity of emerging patterns than previous systems such as GoL. Such observations have made Lenia a particularly interesting system for studying the emergence of life-like phenomenon and even sensori-motor capabilities \citep{hamon2024}. Importantly, these artificial creatures bear a deep similarity to their biological counterparts as enacted agents endowed with constitutive autonomy (i.e creatures showing self-constitution, self-maintenance and some level of behavioral functionality) as described by enactive theories of cognition (see \citet{froese2009} for a more complete discussion on enactive artificial intelligence). \\
However, those patterns are quite difficult to find, necessitating complex fine-tuned search algorithms in order to find update rule parameters allowing the emergence of interesting patterns (e.g spatially localized patterns). Complex and resilient creatures, displaying features of sensorimotor agency, have been found in \cite{hamon2024} using intrinsically motivated goal exploration processes (IMGEP) and gradient descent. \\
Another important challenge in artificial life (ALife) and artificial intelligence (AI) is about the design of systems displaying open-ended intrinsic evolution (i.e unbounded growth of complexity through intrinsic evolutionary processes) \citep{stanley2019}. Such a process is called \emph{intrinsic} since no final objective (i.e fixed fitness function) is set by the experimenter, the fitness landscape is intrinsic to the system and depends only on its current state, as in natural evolution where there is no final goal \citep{lehman2011a}. Seminal works by Von Neumann and Ulam in 1951 have paved the way in this direction. They were particularly interested in building an universal self-reproducing cellular automata (CA) capable of achieving open-ended evolution \citep{vonneumann1966a} quickly followed by Codd's attempt \citep{hutton2010}. Further developments in this direction have quickly followed with Langton's self-replicating loops, a simpler model of self-replication in CA's, at the cost of its universality \citep{langton1984}. Even though, Langton's loops were able to self-replicate, no variations could be introduced in the process, thus making the emergence of evolution impossible. Fifteen years after Langton's self-replicating loops, the goal of obtaining an evolutionary process was achieved by Hiroki Sayama with the Evoloops model which displays Darwinian evolution of self-reproducing Langton's like loops \citep{sayama1999} (see \cite{sayama2024a} for a more complete account of works on evolution and CA). Emergent evolutionary dynamics have also been studied in the context of neural cellular automata \citep{mordvintsev2020,sinapayen2023} and artificial chemistry \citep{kruszewski2022}. \\
However, such systems rely on hand-defined rules, specific structures and controlled settings, ultimately limiting the diversity of patterns that can emerge in the system. On the other hand, even though Lenia creatures display greater diversity, different creatures are governed by different update rules, therefore cannot co-exist in the same world (i.e the same simulation) and cannot interact. Obtaining such an evolutionary process in a CA could be achieved by embedding information in the system locally modifying the update rule and so the properties of emerging creatures, like a genome, enabling multi-species simulations. Such simulations might set the stage for evolution to occur in populations of patterns each with their own update rule and parameters. However, achieving it in CA like Lenia is still an open problem. \\
One very important problem related to this objective is how can one actually measure these emergent evolutionary processes. Two main difficulties exist here. First, in such complex self-organized systems, fitness is intrinsic, i.e there is no externally nor well defined fitness function, thus, one cannot have an objective measure of how adapted an individual is (if there is even a notion of individual). Evolutionary pressures are intrinsic to the system, where self-organised structures have to maintain their own integrity through cooperation or competition with other structures. Secondly, such a measure of evolution should be applicable to a wide range of systems and so rely on as few assumptions about the studied system as possible, fundamental desiderata if one's objective is to study life-as-it-could-be. The framework of evolutionary activity \citep{bedau1996,droop2012} proposed different measures aiming at discerning whether or not evolution is taking place, and quantifying it, in an observed system. Such measures have been applied to artificial systems such as Tierra \citep{ray1999} and Avida \citep{adami1994} and has even been used to compare their dynamics to real-world data \citep{bedau1998}. Importantly, such a measure, or ensemble of measures, could allow to define a clear optimization objective for ALife researchers.\\
We believe that adding mass conservation is a key ingredient to address the aforementioned challenges. Such a constraint could (i) restrict emerging creatures to spatially localized ones, (ii) allow for the design of multi-species simulations and (iii) provide an important evolutionary pressure \citep{hickinbotham2015}. Conservation laws have been thought has fundamental laws for Darwinian evolution to take place \citep{shkliarevsky2019}. We propose in this work a mass-conservative extension to Lenia called \emph{Flow-Lenia} and demonstrate that such conservation laws effectively facilitate the search for artificial creatures by constraining (almost all) emerging patterns to spatially localized ones. We also show that the update rule parameters can easily be optimized using vanilla evolutionary strategies \citep{salimans2017} with respect to some fitness functions to obtain patterns with specific properties such as directed or angular motion. Importantly, we show that the Flow-Lenia formulation enables the integration of the parameters of the CA update rules within the CA dynamics, making them dynamic and localized, allowing for multi-species simulations, with locally coherent update rules that define properties of the emerging creatures. By describing trajectories of parameters over large timescales as well as by using measures of evolutionary activity \citep{bedau1996,droop2012} and diversity, we evaluate the evolutionary dynamics emerging from these multi-species simulations. Moreover, we show that this system is relevant for testing ecological theories of evolution by studying two variations of the vanilla model, one based on dissipative dynamics, and one with resources that creatures need to consume for their survival. The study of the role of dissipative dynamics in this setting is motivated by dissipation having been proposed as one of the four pillars of ``lyfe'', a more general definition of life \citep{bartlett2020}. Introducing dissipative dynamics of the Flow-Lenia system could lead to the emergence of more interesting evolutionary dynamics characterized by higher evolutionary activity measures. On the other hand, resource limitations coupled to a shared pool of resources might create important selective pressures bootstrapping the intrinsic evolutionary process leading to higher evolutionary activity.\\
This paper comes with associated with a companion website \url{\website} showing videos of the system dynamics, as well as open-source code directly executable in an \href{\notebook}{\underline{online notebook}} \footnote{\url{\notebook}}.

\section{Lenia}\label{sec:Lenia}

We here shortly describe the Lenia system, whose dynamics is illustrated Figure~\ref{fig:lenia-flowlenia}.a. For a more detailed explanation, see \cite{chan2019, chan2020, plantec2023b}. Let $\La$ be the support of the CA, here a two-dimensional grid defining the set of cells as well as their spatial relationships. The state of the Lenia system at time $t$ is then defined by the map $A^t: \La \to [0,1]^C$ where $C$ is the number of channels of the system. The system update rule is then defined by the tuple $<K, G, c_1, c_0, A^0>$ where $K$ is a set of convolution kernels with $K_i: \La \to [0, 1]$ satisfies $\int_{\La}K_i = 1$ and $G$ is a set of growth functions with $G_i: [0,1] \to [-1,1]$. Each pair $(K_i, G_i)$ is associated to a source channel $c^i_0$ it senses and a target channel $c^i_1$ it updates. Connectivity can be represented through a square adjacency matrix $M$ of size $C$ where $M_{ij} \in \mathbb{N}$ is the number of kernels sensing channel $i$ and updating channel $j$. $A^0$ is the initial state of the system. As in \citet{hamon2024}, kernels are radially symmetrical and defined as a sum of concentric Gaussian bumps :
\begin{equation} \label{equ:kernel}
        K_i(x) = \sum_{j=1}^{k} b_{i, j} \, exp\left(- \frac{(\frac{x}{r_iR} - a_{i, j})^2}{2 w_{i, j}^2}\right) 
\end{equation}
Where $a_i$, $b_i$, $w_i$ and $r_i$ are parameters defining kernel $i$. $k$ is a parameter defining the number of rings per kernel (set to 3 here) and $R$ is a parameter common to all kernels defining the maximum neighborhood radius. Each kernel is then defined by $3\times k + 1$ parameters.
Growth functions are defined as Gaussian function scaled in the range $[-1, 1]$:
\begin{equation} \label{equ:growth_function}
        G_i(x) = 2 \; exp\left(- \frac{(\mu_i - x)^2}{2\sigma_i^2}\right) - 1
\end{equation}
Where $\mu_i$ and $\sigma_i$ are parameters of growth function $i$ so each growth function is defined by $2$ parameters.
A step in Lenia is defined by the following steps (see figure \ref{fig:lenia-flowlenia} (top)) :
\begin{enumerate}
    \item Compute the growth at time $t$ given the actual state $A^t$ :
        \begin{equation} \label{PotEq}
            U^t_j = \sum_{i=1}^{|K|} h_i \cdot G_i(K_i \ast A_{c^i_0}^t) \cdot [c^i_1 = j]
        \end{equation}
        Where $h \in \R^{|K|}$ is a vector weighting the importance of each pair $(K_i, G_i)$ and $[c^i_1 = j]$ is the Iverson bracket which equals 1 if $c^i_1 = j$ and 0 otherwise (i.e equals 1 if the ith pair updates channel $j$).
        % \begin{equation} \label{PotEq}
        %     U^t_j = \sum_{i} h_{ij} \cdot G_i(K_i \ast A_{c^i_0}^t)
        % \end{equation}
        % Where $h \in \mathbb{R}^{|K| \cdot |K|}$ is a channel adjacency matrix weighting the impact of each pair $(K_i, G_i)$ on the growth
        
    \item Add a small portion of the growth $U^t$ to the actual state $A^t$ to get the state at the next time step and clip results back to the unit range :
        \begin{equation} \label{equ:lenia_update}
            A^{t+dt}_i = [A^t_i + dt \, U^t_i]_0^1
        \end{equation}
\end{enumerate}

\section{Flow-Lenia}\label{sec:FL}

Flow-Lenia extends the Lenia system in the sense that it reuses all the aforementioned components. We propose for this system to interpret activations as concentrations of ``matter'' in all cells and to refer to the term $U^t$, previously called the growth in Lenia, as an affinity map. The idea is that the matter will greedily move towards higher affinity regions by following the local gradient of the affinity map $U$, $\nabla U : \supp \to \R^2$. To do so, we define a flow $F : \mathcal{L} \to (\mathbb{R}^2)^C$, which can be interpreted as the instantaneous speed of matter, as:

\begin{equation}\label{eq:F}
    \begin{cases}
        F^t_i = (1 - \alpha^t) \nabla U^t_i - \alpha^t \nabla A^t_{\Sigma} \\
        \alpha^t(x) = [({A^t_{\Sigma}(x)} / {\beta_A})^n]_0^1
    \end{cases}
\end{equation}

With $A^t_{\Sigma}(x) = \sum_{i=1}^C A^t_i(x)$ the total mass in each location $x$. Here $\nabla U^t_i$ is the affinity gradient for channel $i$. The negative concentration gradient $- \nabla A^t_{\Sigma}$ is a diffusion term to avoid concentrating all the matter in very small regions akin to the clipping in Lenia which upper bounds concentrations. In practice, gradients are estimated through Sobel filtering. The map $\alpha : \mathcal{L} \to [0, 1]$ is used to weight the importance of each term such that $-\nabla A^t_{\Sigma}$ dominates when the total mass at a given location is close to a critical mass $\beta_A \in \mathbb{R}_{>0}$. Intuitively, the result is that matter is mainly driven by concentration gradients in high concentrations regions and is more free to move along the affinity gradient in less concentrated areas. We typically use $n>1$ such that the affinity gradient dominates on a larger range of masses. \\

\begin{figure}
    \centering
        \includegraphics[scale=0.23]{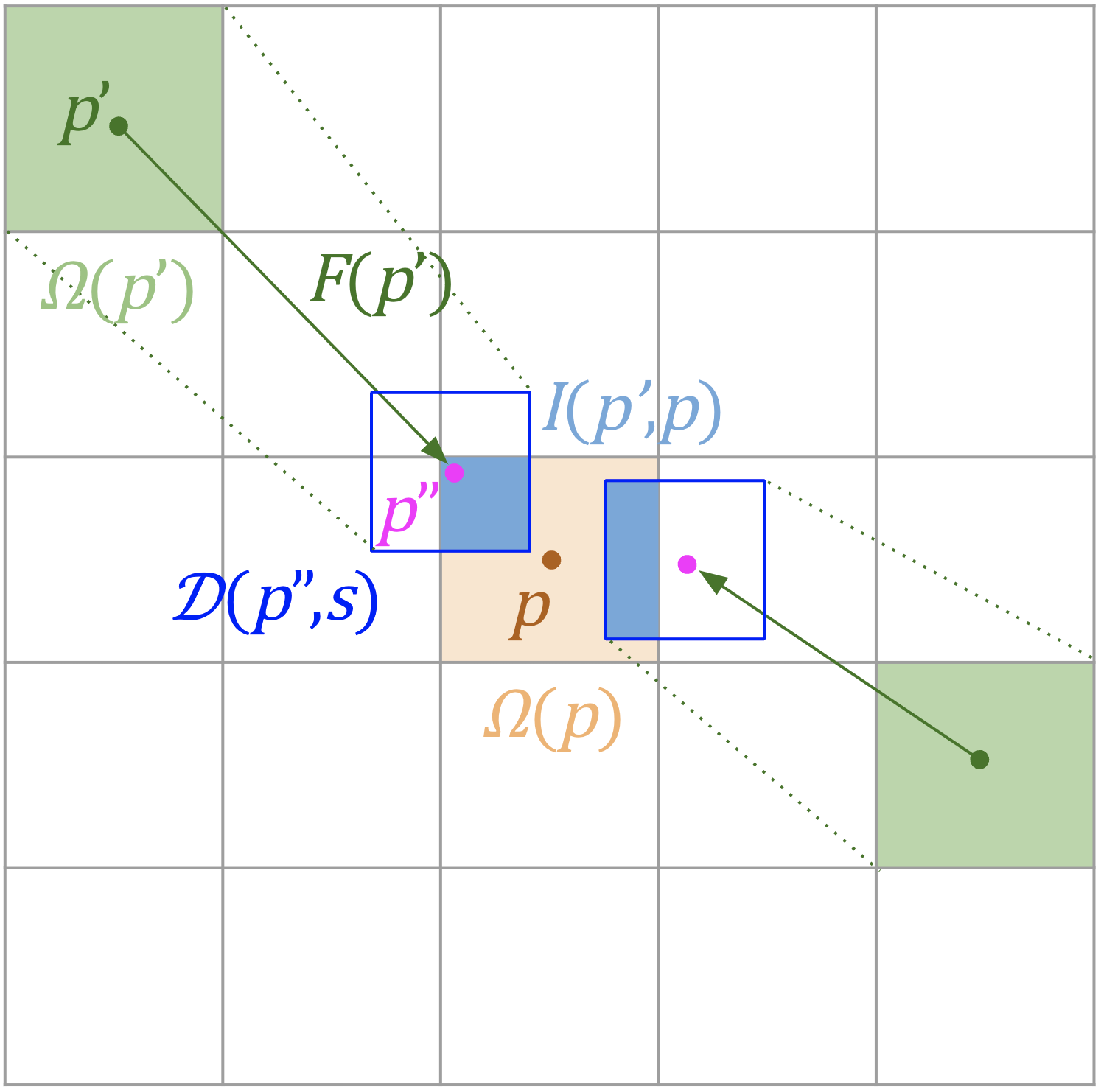}
        \caption{Calculation of incoming matter to cell $p \in \mathcal{L}$ through reintegration tracking \citep{moroz2020}. Mass contained in cell at location $p' \in \mathcal{L}$ is moved to a square distribution $\mathcal{D}$ centered on $p'' = p' + dt \cdot F^t(p')$. The proportion of mass from $p'$ arriving in $p$ is then given by the integral of $\mathcal{D}$ on the cell domain of $p$, $\Omega(p)$, denoted as $I(p', p)$.}
        \label{fig:fig-flow}
\end{figure}

Finally, matter can be displaced in space according to flow $F$ giving us the state at the next time step. To do so we use the reintegration tracking method proposed in \citep{moroz2020}. Reintegration tracking is a semi-Lagrangian grid based algorithm thought as a reformulation of particle tracking in screen space (i.e grid space) aimed at not losing information (i.e particles) which happens when two particles end up in the same cell. The basic principle is to work with distribution of particles (i.e infinite number of particles) and conserve the total mass by adding up masses going on a same cell. Overall, reintegration tracking can be seen as a grid-based approximation to particle systems with infinite number of particles having the property to conserve total mass. Thus, Flow-Lenia can be seen as a new kind of model at the frontier between continuous CA and particle systems. A particle based model directly inspired by the Flow-Lenia formulation has been recently proposed in \citet{mordvintsev2022}. Figure \ref{fig:fig-flow} illustrates how reintegration tracking is used in our case. The resulting update rule (animated in this \href{\video}{\underline{video}} \footnote{\url{\video}}) is the following :

\begin{equation}\label{eq:RT} 
    \begin{cases}
        A^{t+dt}_i(x) = \sum_{x' \in \mathcal{L}} A^{t}_i(x') I_i(x', x) \\
        I_i(x', x) = \int_{\Omega(x)} \mathcal{D}(x_i'', s) 
    \end{cases}
\end{equation}

With $x_i'' = x' + dt \cdot F^t_i(x')$ the target location of the flow from $x'$ in channel $i$. $\Omega(x)$ is the domain of cell at location $p$, which is a square of side $1$. $\mathcal{D}(m, s)$ is a distribution defined on $\mathcal{L}$ with mean $m$ and variance $s$ satisfying $\int_{\mathcal{L}} \mathcal{D}(m, s) = 1$, which is in practice a uniform square distribution with side length $2s$ centered at $m$. This distribution emulates a flow of particles from source area $\Omega(x')$ to target area $\mathcal{D}(x'', s)$, where the distribution $\mathcal{D}$ emulates Brownian motion at the low level. $s$ is an hyperparameter of the system which can be seen as form of temperature. The reintegration tracking method is depicted in Fig. \ref{fig:fig-flow}. Since the distribution $\mathcal{D}$ integrates to 1, it is clear that a cell cannot send out more mass than it contains nor less and so the system conserves its total mass. For computational reasons, we do not look at all cells to compute incoming matter as described by equation \ref{eq:RT} but only at the neighborhood composed of cells whose Chebyshev distance to the target cell is less than 5 (extended Moore neighborhood) allowing for considerably reduced computation times. Mass conservation also implies that cells' states are no longer bound to the unit range but can be any positive real valued number ($S \equiv \mathbb{R}_{\geq 0}^C$). This model has been implemented in JAX \citep{bradbury2018} allowing fast simulation on GPU ($255 \mu s \pm 3.11 \mu s$ per step on Tesla T4 GPU with 1 channel, 10 kernels and $128 \times 128$ world size).

\subsection{Flow-Lenia with parameters embedding}\label{sec:FLP}

Flow-Lenia formulation, by considering a flow of matter, allows to attach any information to the moving matter such as the update rule parameters making them dynamic and localized. Formally, this comes to define a parameter map $P: \La \to \Theta$ where $\Theta$ is the parameter space. In this work, only the kernel weighting vectors $h$ are included in the parameter space $\Theta \equiv \mathbb{R}^{|K|}$. This map can then be used to compute the affinity score in each cell $x$ by weighting the influence of each pair $(K_i, G_i)$ with the localized vector $P(x)$ giving the following formula:

\begin{equation}\label{eq:UP}
    U^t_j(x) = \sum_{i=1}^{|K|} P^t_i(x) \cdot G_i(K_i \ast A_{c^i_0}^t)(x) \cdot [c^i_1 = j]
\end{equation}

While in theory, all the parameters could be embedded in the parameter map, this would come with high memory and computational costs for some. In particular, changing the kernels parameters dynamically would make the use of fast convolution operations such as fast-Fourier convolution impossible as it would require using different kernels in all different locations of the map.\\
We can now move the parameters along with the matter during the reintegration tracking phase. This necessitates deciding what to do when different set of parameters arrive in a same cell. Different versions of the mixing rule have been proposed in \cite{plantec2023b} either based on averaging incoming parameters or sampling one based on quantities of incoming matter. In this work, we use the latter one which is defined by the following equation for the stochastic version:

\begin{equation}\label{eq:mix}
    \mathbb{P}[P^{t+dt}(x) = P^{t}(x')] = \frac{e^{A^t(x') I(x', x)}}{\sum_{x'' \in \mathcal{L}}e^{ A^t(x'') I(x'', x)}} 
\end{equation}

This comes to sampling one of the parameters where probabilities for each cell to be selected are given by the softmax distribution computed from associated incoming quantities of matter. We also used a deterministic version of this rule where the softmax sampling is simply replaced by an argmax. However, when not stated otherwise, the method we use is the stochastic sampling. We chose this rule because of the competitive dynamics it creates in the system. It enables creatures to convert mass from other creatures with different parameters (i.e other species). This would not be possible with the average rule as each interaction would create a new set of parameters.

\section{Experimental methods}

The experiments are divided into three main parts. First, we perform random search in the Flow-Lenia parameter space allowing us to qualitatively analyze the dynamics of the system and the typical patterns emerging from it. In a second part, we optimize the update rules parameters as well as the initial pattern configuration in order to obtain creatures displaying specific behaviors. Finally, we experiment with the parameters embedding mechanism and analyze the long-term temporal dynamics emerging from these multi-species simulations. Each of these experiments are explained in detail in sections \ref{meth:rs}, \ref{meth:ds} and \ref{meth:evo} respectively, and the associated results are presented in section~\ref{sec:results}. Methods and results from random and directed search experiments are also presented in more details in \cite{plantec2023b}.

\subsection{Random search experiments}\label{meth:rs}

We performed random search in the Flow-Lenia parameter space described in table \ref{tab:flow_params}. We refer the reader to sections \ref{sec:Lenia} and \ref{sec:FL} for further details on the role of these parameters. Associated results are presented in section \ref{sec:RS}.

\begin{table}[ht]
    \centering
    \begin{tabular}{|c c c | c c c|}
        \hline
        \multicolumn{3}{|c|}{\textbf{Neighborhood}} & \multicolumn{3}{|c|}{\textbf{Growth functions}}\\
        \hline
        $R$ &  $\in [2, 25]$ & & $\mu$ &   $\in [0.05, 0.5]$ & *\\
        $r$ &  $\in [0.2, 1]$ & * & $\sigma$ &  $\in [0.001, 0.2]$ & * \\
        \hline 
        \multicolumn{3}{|c|}{\textbf{Kernels }} & \multicolumn{3}{|c|}{\textbf{Flow }}\\
        \hline
        $h$ &  $\in [0, 1]$ & * & $s$ &  $0.65$ & \\
        $a$ &  $\in [0, 1]^3$ & * & $n$ &  $2$ & \\
        $b$ &  $ \in [0, 1]^3$ & * & $dt$ &  $0.2$ & \\
        $w$ &  $\in [0.01, 0.5]^3$ & * & & & \\
        \hline 
    \end{tabular}
    \caption{Flow Lenia explored parameter space. Parameters marked with a * must be sampled for each kernel-growth function pair.}
    \label{tab:flow_params}
\end{table}

\subsection{Directed search experiments}\label{meth:ds}

We used evolutionary strategies \citep{salimans2017} to optimize the update rule parameters and the initial configuration $A^0$. We trained the model with respect to four different user-defined fitness functions, i.e tasks: directed motion, angular motion, navigation through obstacles and chemotaxis. We refer the reader to \citet{plantec2023b} for further details about the employed fitness functions.
We used EvoSax \citep{lange2022} implementation of the OpenES \citep{salimans2017} strategy with population size of $16$ and Adam optimizer \citep{kingma2017} with $0.01$ as learning rate. We optimized the Flow Lenia update rule with different number of kernels and either 1 or 2 channels. For comparison, we also trained original Lenia on the directed motion task following the same optimization procedure. The initial pattern is composed, as in random search, of a square patch with non-zero activations placed at the center of the world and zeros everywhere else.

\subsection{Intrinsic evolution experiments}\label{meth:evo}

In order to analyze the potentially evolutionary dynamics enabled by the parameters embedding mechanism presented in section \ref{sec:FLP}, we performed simulations with larger spatial and temporal scales. A similar attempt using the original Lenia system for large-scale simulation of intrinsic evolution was described in \citep{chan2023}. By allowing multispecies simulations, the parameters embedding mechanism also allows for interspecies competition especially under the stochastic parameter selection rule described in equations \ref{eq:UP} as it allows species to convert matter from other species. We further propose two variations of the vanilla model, namely the dissipative and food models. During simulation, the set of unique parameters denoted $\mathcal{P}^t\equiv\{P^t(x)\}_{x\in \La}$ is recorded. Together with parameters, we record their associated total mass through time $M(p,t)=\sum_{x\in\La}A_{\Sigma}^t(x)\cdot[P^t(x)=p]$. We also introduce a diversity metric $D(t)$ quantifying the diversity of parameters and which is defined as the average distance between all present parameters in the system at this time $t$:

\begin{equation}\label{eq:D}
    D(t) = \frac{1}{|\mathcal{P}^t|} \sum_{p \in \mathcal{P}^t} \sum_{p' \in \mathcal{P}^t} ||p-p'||_2
\end{equation}

Where $||\cdot||_2$ is the euclidean norm. 
We only recorded simulation data every 100 steps for memory reasons. However, as interesting creatures' behaviors unroll in around 100 time steps, evolutionary dynamics must happen on much larger scales. The simulation settings as well as the three different models are described in more details in the following sections.

\subsubsection{Simulation settings}

All presented models have been simulated for $500 \cdot 10^3$ steps. The system is initialized with, when not stated otherwise, 3 channels and 5 kernels per channel pair making a total of 45 kernels. We introduce mutations in the form of square ``beams'' affecting a random $10\times 10$ patch in the grid. Beams apply a perturbation sampled from a normal distribution with mean $0$ and unit variance to the parameter map $P$, the perturbation being the same for all cells in the affected patch. While we could have implemented mutations on single cells only, it would have been unlikely for any mutation to have the opportunity to develop as they would be quickly overtaken by their neighbors. Affecting larger zones using beams gives a mutation better chances of developing. Mutation rates are controlled by the parameter $p_{mut}$ which is the probability of a mutation beam appearing at each time step. All simulations have been repeated with $5$ different random seeds. 

\subsubsection{Model variations} \label{sec:models}

We propose in this work three different variations of the Flow-Lenia model namely: vanilla, dissipative and food which are presented here after. 

\paragraph{Vanilla}

In this setting, the environments is simply initialized with 64 creatures. Each creature is initialized as a $20 \times 20$ square patch which position is uniformly sampled on the grid $\La$. Matter concentrations ($A$) in these patches are also sampled uniformly in $[0,1]$ and parameter ($P$) is sampled following a normal distribution and set identically for all cells in a patch. \\
In this setting, high fitness parameters are the ones leading to creatures able to preserve and increase their associated mass. Note that here, creatures can only grow by converting matter from other creatures. This creates pressures for strong individuality, especially with the stochastic update rule, as it incentives creatures to protect their resources. But strategies that are too defensive, preventing a parameter set from expanding (i.e gaining mass and territory), might put it at risk as it might make it more vulnerable to disappear because of a random mutation beam.

\paragraph{Dissipative}

In the dissipative setting, the world is initialized in the same way as the vanilla setting and mutations are also used in the same way. The difference is that in this setting we regularly remove matter and the associated parameters and add new ones, thus creating dissipative dynamics. To do so, we use two new types of beams, one removes matter and parameters in the affected patch, the other adds a new creature (i.e a patch with random parameters, as in the initial pattern) at another affected location sampled in a $100\times 100$ corner of the grid, the input zone, with randomly initialized parameters. The new creature is initialized in the same way as the initialization phase. The rate at which the dissipative beams appear is controlled by parameter $p_{diss}$. \\
We expect the dissipative setting to create more interesting evolutionary dynamics characterized by higher evolutionary activity measures (see section \ref{sec:MEA})  by creating an environment with a constant input of novelty in the form of new parameters  while conserving more stable zones in the environments (i.e the ones further from the input zone).

\paragraph{Food}\label{food_model}

In this last setting, we use the food mechanism displayed in figure \ref{fig:lenia-flowlenia} and presented in \cite{plantec2023b}. Creatures see their mass constantly decaying at a fixed rate $r_{decay}$. We then introduce a new map $\Psi: \La \to [0, \infty)$ defining the presence of food for each cell in the world. We also add kernels sensing the food map enabling creatures to sense food. When matter and food are present in a same cell, then food is going to be transformed into matter at a rate $r_{digest}$. The food map is initialized with 32 $5\times 5$ food squares (where value of the map is set to 1 for all cells in the patch) randomly sampled on the grid. At each time step, a new food patch is added with probability $p_{food}$. \\
In this model, a high fitness creature is one able to counter its constant decay by either converting matter from other creatures (i.e being a predator), or consuming food resources. Such a constraint for creatures, creating a need to find food for their continued existence, together with a common pool of resources, might create strong evolutionary pressures and competitive dynamics, bootstrapping the emergence of evolutionary processes in the system. Moreover, the addition of the food constraint and the necessity to counter-act their decay introduce the notion of a minimal criterion, i.e a criterion creatures must met in order to expand, which has been proposed as a fundamental ingredient for open-ended evolution to emerge \citep{soros2014,taylor2015a}. 

\subsubsection{Measuring evolutionary activity}\label{sec:MEA}

Multiple measures of evolutionary activity have been proposed in literature. In this work, we use two different measures, namely count-based evolutionary activity ($EA^C$) and non-neutral evolutionary activity ($EA^N$) \citep{droop2012}. Evolutionary activity metrics are based on records of the presence and counts of different components in a system. Components can be for instance molecules or species. In our case, the components are the different parameters, meaning that a component, or species, is a unique point in the parameter space. Hence, two set of parameters with infinitely small differences are considered as different species, regardless of their phenotypic outcome. This is a current limitation of the study as discussed in the section \ref{sec:disc}. We here use both parameters sets $\mathcal{P}^t$ and associated masses $M$ to compute component level activities $a^C_p(t)$ and $a^N_p(t)$ for count-based and non-neutral activities respectively. Count-based activity is based on the total mass associated to a given set of parameters. At each time step, the count-based activity of a component is incremented by its total associated mass if this component exists:

% \begin{equation}
%     a^C_p(t) = \begin{cases}
%         a^C_p(t-1) + M(p,t) \textrm{ if } p \in \mathcal{P}^t \\
%         0 \textrm{ else}
%     \end{cases}
% \end{equation}

\begin{equation}
    a^C_p(t) = (a^C_p(t-1) + M(p,t)) \cdot [p \in \mathcal{P}^t]   
\end{equation}

Where $[x]$ is the Iverson bracket which equals $1$ if $x$ is true and $M(p,t)$ is the mass associated to species $p$ at time $t$.\\
Intuitively, the greater the total mass of a parameter $p$ is, and the longer it survives, the higher will be its activity $a_p$. While count-based activity can give useful insights, it does not tell much about the quantity of change happening in the environment. Non-neutral activity solves this issue by penalizing periods of stasis. Here, the activity of a component is incremented by the square of the change of its proportion in the population of components if it increased. Hence, if multiple components stay very stable, i.e reach a stable equilibrium, their activities will remain constant. However, if a component sees its proportion in the population going up then its activity will increase. This measure thus prevent periods of stasis from contributing to the evolutionary activity measure. This is formally defined by the following set of equations:

\begin{equation}
    \begin{cases}
        % a^N_p(t) = \begin{cases}
        %     a^N_p(t-1) + \Delta^N_p(p,t) \textrm{ if } p \in \mathcal{P}^t \\
        %     0 \textrm{ else}
        % \end{cases}\\
        a^N_p(t) = (a^N_p(t-1) + \Delta^N_p(p,t)) \cdot [p \in \mathcal{P}^t]\\
        \Delta^N_p(t) = (\sum_{p'} M(p',t)) \cdot (\rho(p,t)-\rho(p,t-1))^2 \cdot [\rho(p,t) > \rho(p, t-1)]\\
        \rho(p,t) = \frac{M(p,t)}{\sum_{p'}M(p',t)}
    \end{cases}
\end{equation}

Global activity at a given time is simply defined as the sum of all components activities at this same time step: $EA^*(t)=\sum_{p}(a^*_p(t))$ where $*$ is either $C$ or $N$ for count-based and non-neutral evolutionary activities respectively. 

\section{Results}
\label{sec:results}

\subsection{Random search}\label{sec:RS}

\begin{figure}[ht!]
    \subfloat[Flow-Lenia]{\includegraphics[width=.9\textwidth]{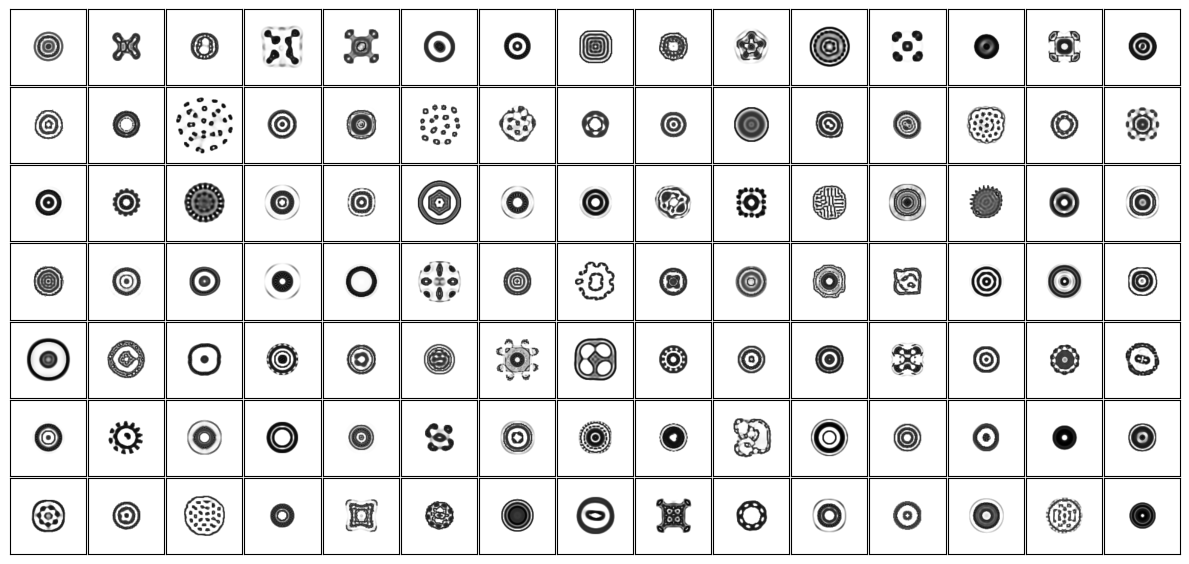}} \\
    \subfloat[Lenia]{\includegraphics[width=.9\textwidth]{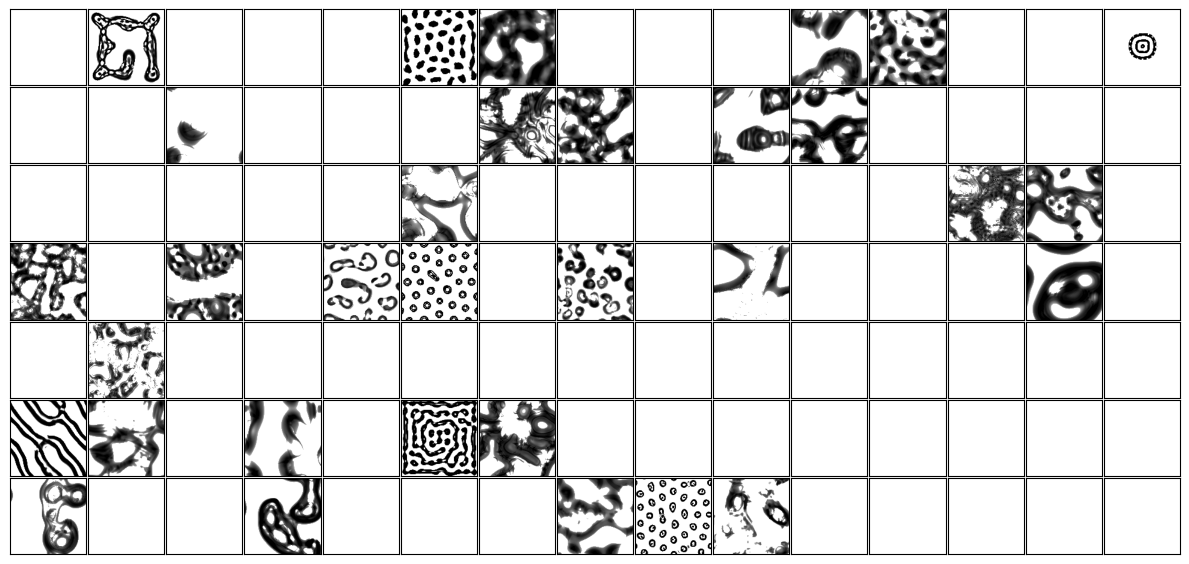}} \\
\caption{Patterns obtained from $105$ different randomly sampled update rule parameters in (a) FLow-Lenia and (b) Lenia systems. Each pattern is obtained by simulating the systems for $150$ steps from an initial state composed of a $40\times 40$ patch with uniformly sampled concentrations. The exact same $105$ parameter sets are used for both systems.}
\label{fig:RS2}
\end{figure}

\begin{figure}[ht!]
    \begin{tabular}{ccc}
         \subfloat[]{\includegraphics[width=.2\textwidth]{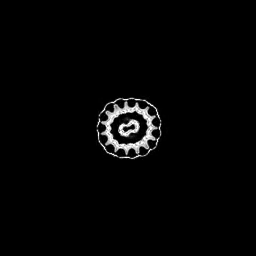}}&  
         \subfloat[]{\includegraphics[width=.2\textwidth]{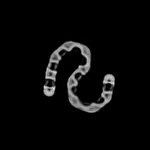}}
         &
         \subfloat[]{\includegraphics[width=.2\textwidth]{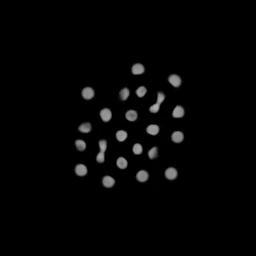}}\\
         \subfloat[]{\includegraphics[width=.2\textwidth]{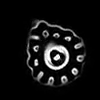}}
         &  
         \subfloat[]{\includegraphics[width=.2\textwidth]{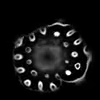}}
         &
         \subfloat[]{\includegraphics[width=.2\textwidth]{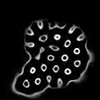}}\\
         \subfloat[]{\includegraphics[width=.2\textwidth]{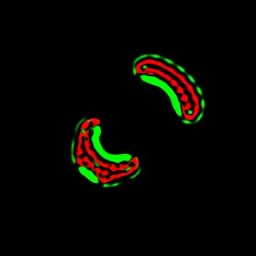}}&  
         \subfloat[]{\includegraphics[width=.2\textwidth]{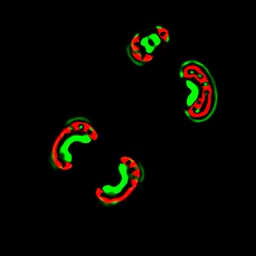}}
         &
         \subfloat[]{\includegraphics[width=.2\textwidth]{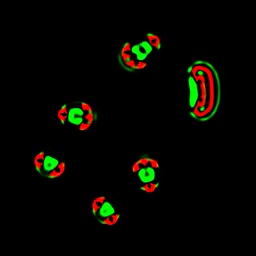}}\\
         \multicolumn{3}{c}{\subfloat[]{\includegraphics[width=.65\textwidth]{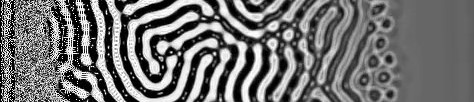}}}
    \end{tabular} 
\caption{Flow Lenia creatures. (a-c) Samples of creatures found through random search in Flow Lenia parameter space. (d-f) and (g-i) Timelapses of patterns found through random search. Colors in (g-i) code for different channels. (j) Effect of changing temperature in Flow Lenia, temperature is linearly increasing from left to right. Videos are available at \url{\website}.}
\label{fig:RS}
\end{figure}

By performing random and manual search of the Flow Lenia parameter and hyperparameter space described in table \ref{tab:flow_params} we have been able to discover SLPs with already interesting and complex behaviors some of which are displayed in figures \ref{fig:RS2} and \ref{fig:RS}.  \\
Most of the patterns generated in Flow Lenia are SLPs (see figure \ref{fig:RS2}(a)) with rare exceptions found by manually setting parameters to specific configurations leading to scattered matter. We can see in figure \ref{fig:RS}(b) that the same parameters mostly lead to empty or exploding patterns in the Lenia system. Using multiple kernels led to the emergence of SLPs with more complex shapes and behaviors. While part of emerging patterns tend to be static ones, dynamic patterns are quite common in Flow Lenia. For instance gyrating SLPs (Fig.~\ref{fig:RS}.a) or snake like patterns (b) with complex motion emerging from attraction/repulsion dynamics can be frequently observed. Dividing and merging dots (c) resembling reaction-diffusion patterns are also a common pattern. Timelapse (d-f) shows a creature with complex and unpredictable dynamics emerging from the interactions of its membrane, multiple organoids-like structures and a central nuclei ultimately leading to a phase transition happening in (e). Timelapse (g-i) shows a 2-channels creature displaying complex division patterns and interesting modular creatures whose characteristics change depending on their total mass while being of the same ``kind'' (i) (see 5 creatures on the leftmost part of (i)). Note that multi-channel creatures often show more complex dynamics and patterns with very modular shapes where each channel seems to occupy a different role. (j) shows the effect of changing the size of the reintegration tracking distribution $s$ (see equation \ref{eq:RT} and figure \ref{fig:fig-flow}), parameter we call temperature. Here temperature is linearly increasing from left to right showing very different phases of the systems. More interestingly, patterns at the frontier between the Turing-like phase (center) and the equilibrium phase (right) are much more dynamic and display unpredictable dynamics suggesting a critical regime. 

\subsection{Optimizing Flow Lenia creatures}\label{sec:DS}

\begin{figure}[t]
\centering
\begin{tabular}{cc}
    \begin{tabular}{c}
         \subfloat[]{\includegraphics[width=.5\textwidth]{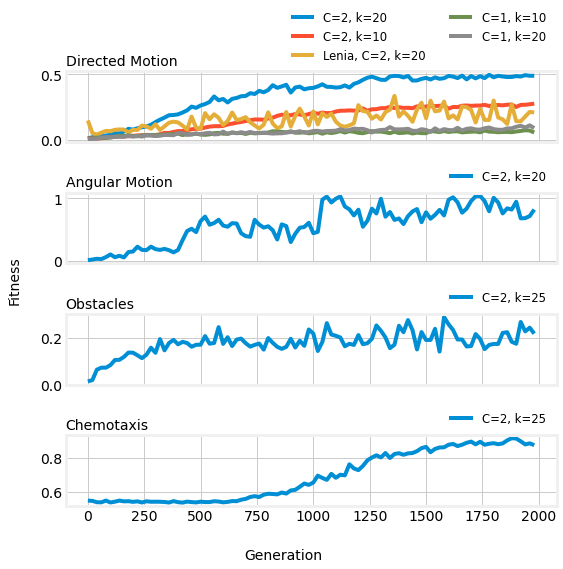}}
    \end{tabular}
    & 
    \begin{tabular}{cc}
         \subfloat[]{\includegraphics[width=.2\textwidth]{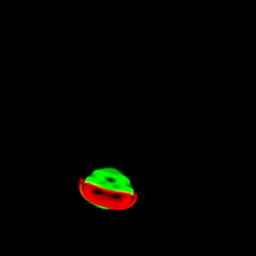}} &
         \subfloat[]{\includegraphics[width=.2\textwidth]{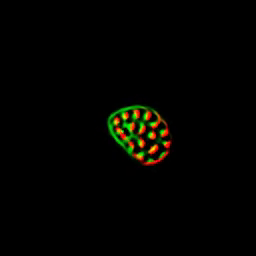}} \\
         \subfloat[]{\includegraphics[width=.2\textwidth]{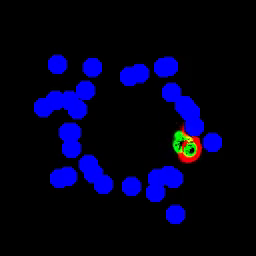}} &
         \subfloat[]{\includegraphics[width=.2\textwidth]{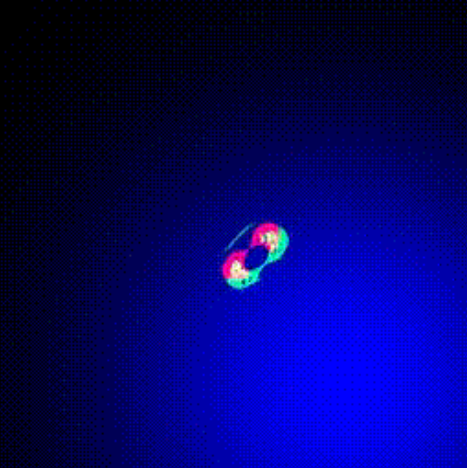}} \\
    \end{tabular}
\end{tabular}
\caption{(a) Results of evolutionary optimization. C is the number of channels of the system and $k$ is the number of kernels and growth functions. When performing the exact same optimization for directed motion in the original Lenia system (yellow curve), not only optimization is unstable but it only discovers exploding patterns. (b-e) Creatures found through optimization. (b) Directed motion with 2 channels and 20 kernels. (c) Angular motion with 2 channels and 20 kernels. (d) Motion through obstacles with 2 channels and 25 kernels. (e) Chemotaxis with 2 channels and 25 kernels. Videos are availabe at \url{\website}.}
\label{fig:evo_creas}
\end{figure}

Flow-Lenia update rule parameters can also be easily optimized so to generate patterns with specific behaviors. This is a difficult task in Lenia as it would require constantly monitoring the existential status and the spatially-localizedness of evolved creatures. Thus training creatures in Lenia requires to define characterizations of creatures accounting for such properties which is a far from trivial problem. Moreover, even if one can come up with proxies to find spatially localized patterns, the optimization process remains difficult necessitating advanced optimization methods like curriculum learning \citep{hamon2024}. In Flow Lenia, the spatial localization constraint is intrinsic to the system thus removing the necessity to account for it when searching for creatures.  \\
Using evolutionary strategies \citep{salimans2017} we have been able to find creatures solving various tasks such as:

\paragraph{Directed motion} 
The creature is able to move as fast as possible in one direction. Efficient solutions can be found in the 2 channels condition but not in the single channel case. However, when running the optimization algorithm for longer (e.g 5000 generations), we have been able to find single channel creatures with similar fitness than their 2 channels counterpart. Increasing the number of kernels led to faster discovery of good solutions. The best performing creature is shown in figure \ref{fig:evo_creas}(b). This creature moves because of attraction/repulsion dynamics between the 2 channels which might explain why directed motion is much easier to attain with multi-channels creatures. On the other hand, the optimization of the original Lenia model is much less stable and discovered patterns are less successful than their mass-conservative counterparts. Moreover, every Lenia optimized patterns are exploding ones.

\paragraph{Angular motion} 
The creature is able to maximize its straight line speed as well as to make turns. The best performing creature, shown in figure \ref{fig:evo_creas} (c), displays very complex internal dynamics leading it to periodically make 180° turns while moving in straight line the rest of the time. These dynamics seem to be generated by attraction repulsion dynamics like the ones observed in directed motion but here in a more intricate morphology.

\paragraph{Navigation through obstacles} 
The creature is able to maximize its traveled distance while multiple obstacles are placed on its way. We have been able to successfully train creatures able to move and maintain their integrity when making contact with walls such as the one shown in figure \ref{fig:evo_creas} (d) which is able to resist deformation and find a way out the ``forest". In comparison, solving a similar task in Lenia required complex optimization methods based on curriculum learning, diversity search and gradient descent over a differentiable CA \citep{hamon2024}. However, such a comparison is difficult because Flow Lenia creatures are inherently more robust due to conservation of mass, whereas Lenia creatures can disappear because of perturbations.

\paragraph{Chemotaxis} 
The creature is able to follow a concentration gradient, encoded in a separate channel and which it is able to sense through some kernels, towards its source. The best solutions such as the one shown in figure \ref{fig:evo_creas} (e) are perfectly able to climb the gradient towards its maximum.

For further details on the optimization procedure, we refer the reader to \cite{plantec2023b}.

\subsection{Intrinsic evolutionary dynamics}\label{sec:Evo}

\begin{figure}
     \centering
     \subfloat[][Vanilla]{\includegraphics[width=.7\textwidth]{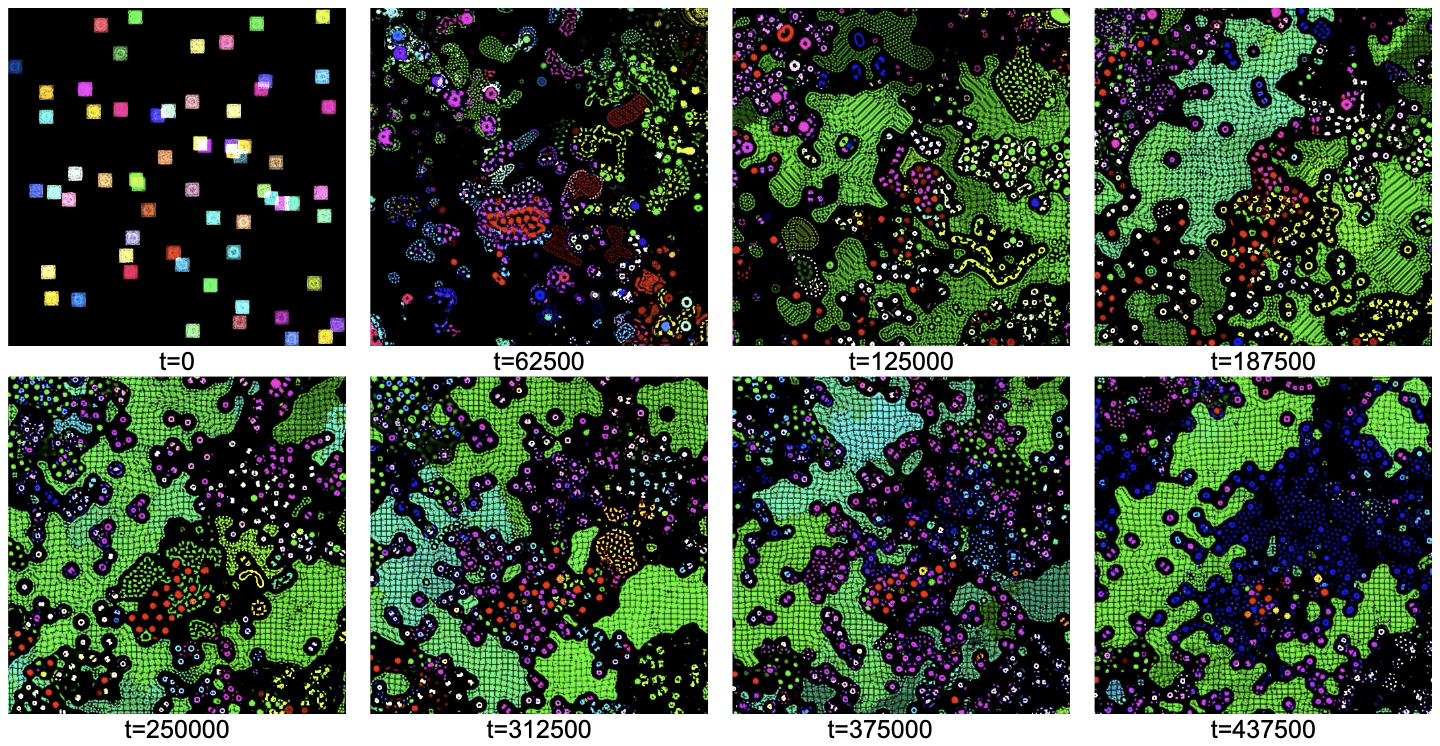}}\\
     \subfloat[][Dissipative]{\includegraphics[width=.7\textwidth]{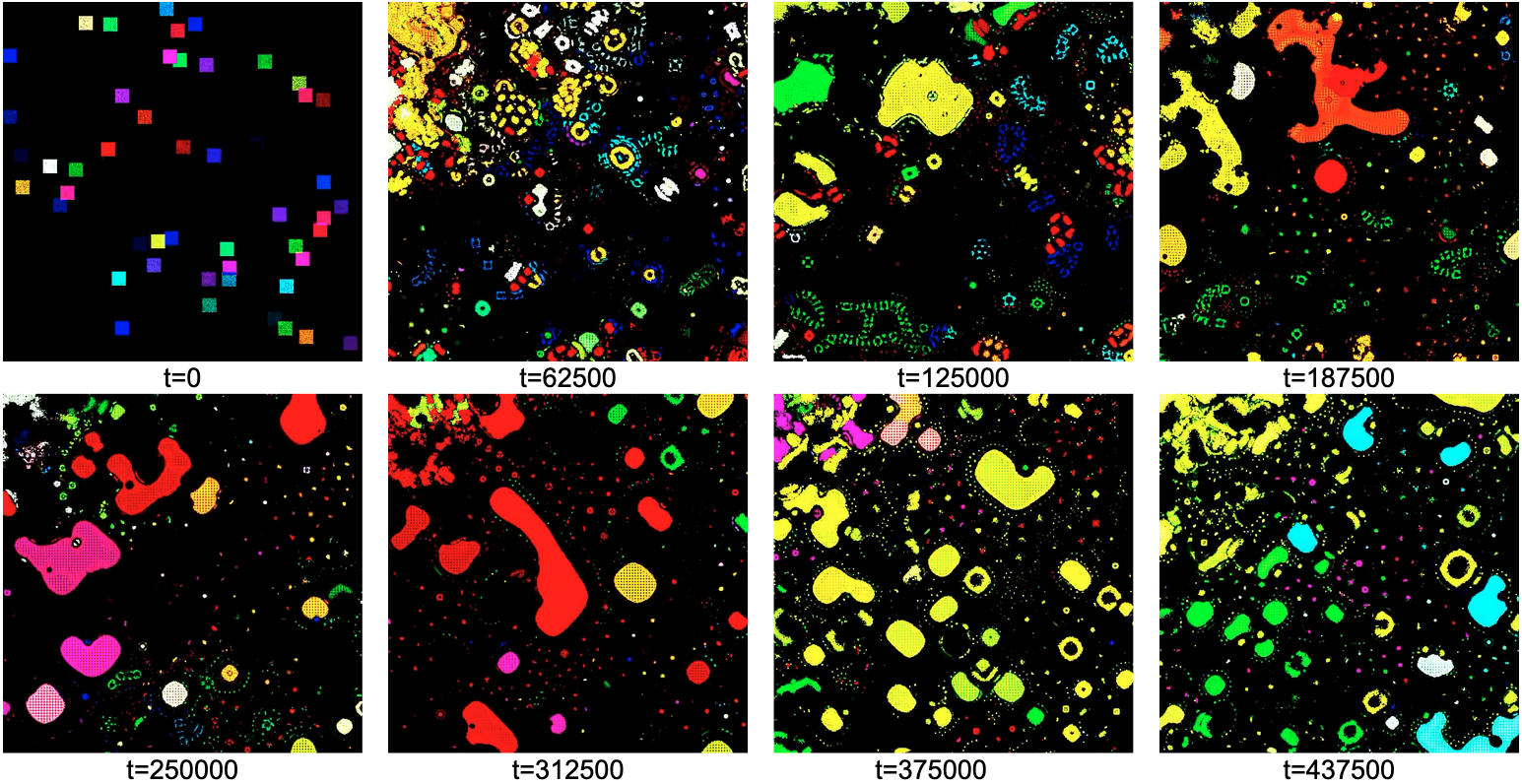}}\\
     \subfloat[][Food]{\includegraphics[width=.7\textwidth]{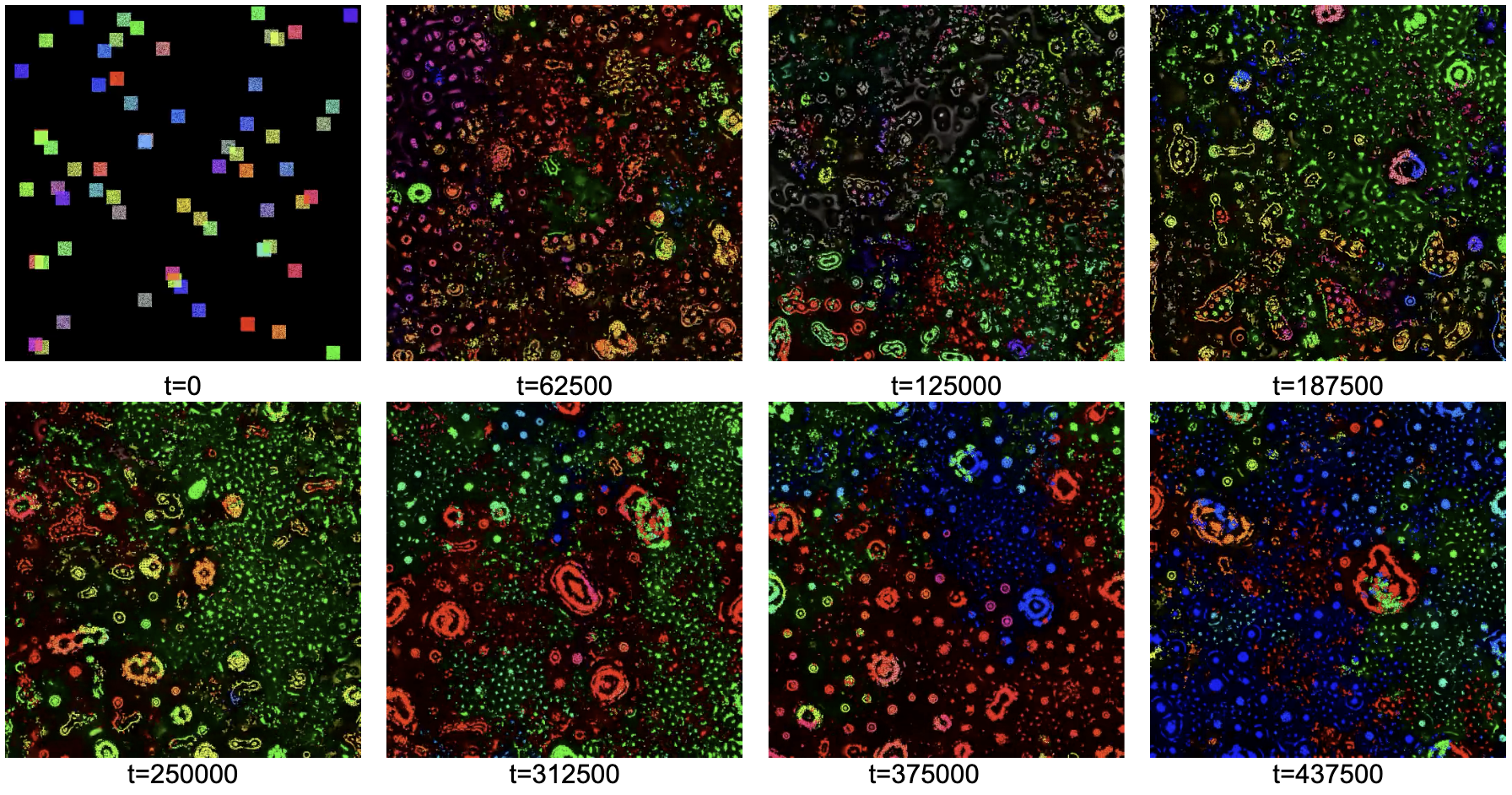}}
     \caption{Snapshots of simulations for the vanilla (a), dissipative (b) and food model (c). Colors are defined by the parameter map while intensity is set by concentrations of matter. (a) shows a very large and stable green structure, instance of a larger scale creature. Videos of different simulations are available in the associated website \url{\website}}
     \label{fig:snaps}
\end{figure}

\begin{figure}[tp] % Example Figure
    \centering
    \subfloat[]{\includegraphics[width=.4\textwidth]{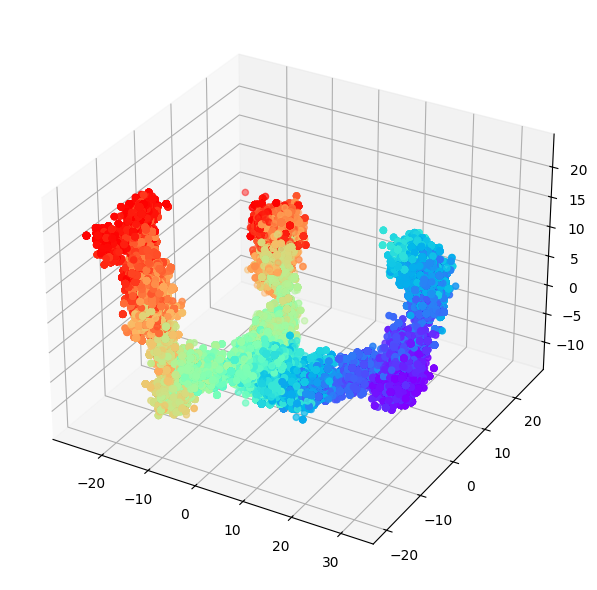}}\subfloat[]{\includegraphics[width=.4\textwidth]{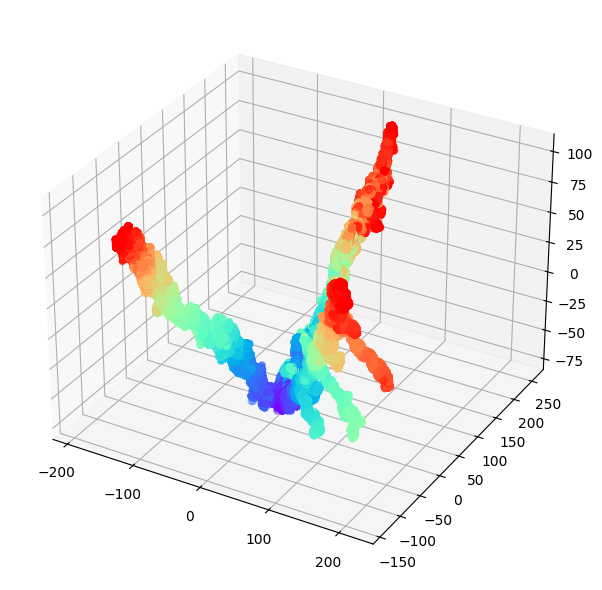}}\\
    \subfloat[]{\includegraphics[width=.4\textwidth]{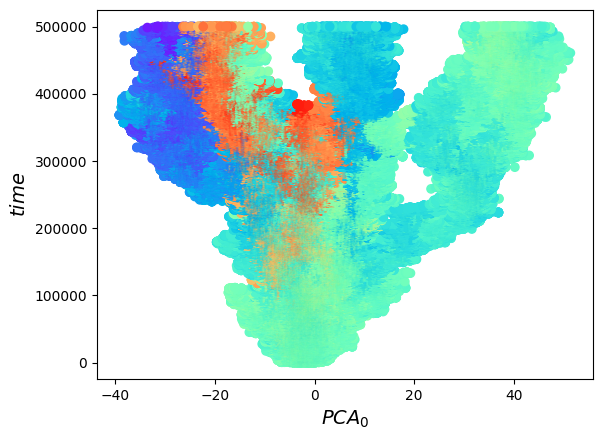}}
    \caption{Visualization of Flow-Lenia evolutionary trajectories through projection in the parameter subspace formed by the principal components of the set $\mathcal{P}$ of parameters having existed during the simulation. (a) and (b) show two different evolutionary trees obtained from simulations of the food model (a) and vanilla model with deterministic parameter mixing rule (b). Colors are coding for time (from purple to red) showing for instance two branches having survived and one which went extinct in (a). (c) is an alternative visualization of the data in 2 dimensions where the x axis is the first principal component, the y axis is the time axis, colors code for the second principal component obtained from a simulation with the vanilla model.}
    \label{fig:evo_tree}
\end{figure}

\begin{figure}[tp] % Example Figure
    \centering
    \includegraphics[width=1.\textwidth]{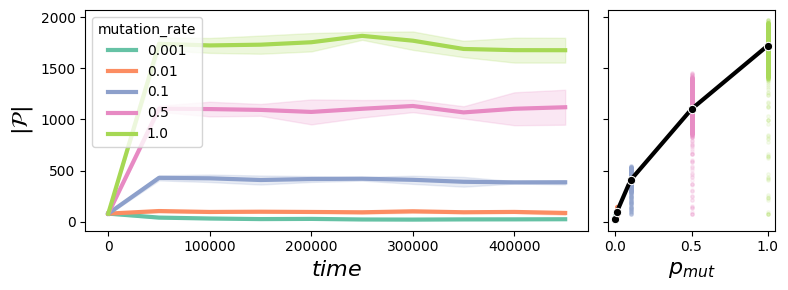}
    \caption{Evolution of the number of different parameters through time for different mutation rates $p_{mut}$. (left) The number of different parameters is plotted against time where error bands correspond to the standard deviation over 5 different runs. (right) The average number of different parameters over time is plotted against $p_{mut}$.}
    \label{fig:P_mut}
\end{figure}

\begin{figure}[tp] % Example Figure
    \centering
    \includegraphics[width=1.\textwidth]{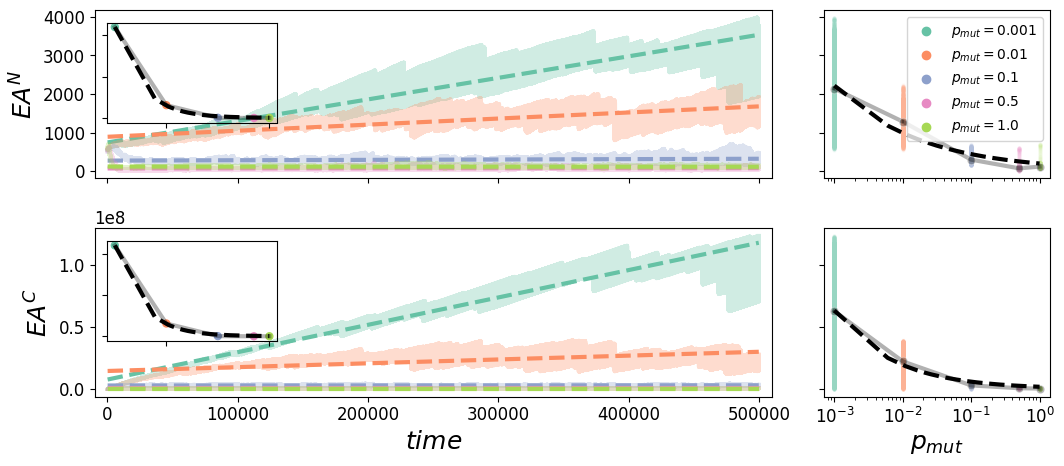}
    \caption{Evolutionary activity measures through time for different mutation rates $p_{mut}$ in the vanilla model. Intuitively, the greater the total mass of a parameter $p$ is, and the longer it survives, the higher will be its count-based activity ; with an additional penalization for stasis for the non-neutral activity (see section \ref{sec:MEA} for formal definitions). Measures of non-neutral (top) and count-based evolutionary activity (bottom) are shown. (left) Evolution of the evolutionary activity is shown as a function of time, dotted lines show the best fitting linear model. The inner plots show the the slopes of these models against their respective mutation rates $p_{mut}$, dashed lines indicate best fitting power function. (right) Average values of evolutionary activity over time are plotted against mutation rates, dashed line shows the best fitting power function.}
    \label{fig:ea_mut}
\end{figure}

Snapshots of sampled simulations can be seen in figure \ref{fig:snaps}. In all settings we tried, we visually observed changes in the set of parameters present in the environment through time, often with some species taking over others, leading to extinctions, and mutations giving rise to new species able to survive. We also observe interesting interspecies dynamics where different parameters form stable structures without competing which can be seen as some form of cooperation or symbiosis. It should be noted that the general appearance of simulations highly depends on the parameters of the system, i.e the kernels and growth function parameters which stay fixed during simulation. While some simulations are visually appealing to us, displaying creatures with different scales and behaviors, other might look much more chaotic with only very small dot-like creatures creating poorly human-readable dynamics. \\
In order to analyse how the parameters evolve through time and move in the parameter space, we first obtained the subspace of maximal variation of the parameter space through Principal Component Analysis (PCA). Principal Components were fitted with the complete set of parameters having existed through simulation $\mathcal{P} \equiv \cup_{0\leq t \leq T} \mathcal{P}^t$, where $\mathcal{P}^t$ is the set of parameters present in the world at time $t$. Then, we can visualize trajectories in the parameter space by projecting the set of parameters present at each timestep $\mathcal{P}^t$ in this subspace. Example of a trajectories are shown in figure \ref{fig:evo_tree}. Interestingly, we can notice that motion in this space looks far from random and take the form of a tree, an evolutionary tree. We can see the formation of different branches which could be seen as instances of speciation. This differentiation clearly indicates the presence of an intrinsic fitness landscape. We also observe that trees obtained from simulations with the deterministic sampling rule for parameters (see sections \ref{sec:FLP}) have much thinner branches and distinctive trajectories. Interestingly, the stochastic sampling rule only adds more stochasticity in the interactions between species, not on the parameters, indicating that these trajectories are effectively produced by an intrinsic fitness landscape seemingly sharper under the deterministic sampling rule (i.e with less noise in the intrinsic fitness of a creature). It is important to note that these dynamics are clearer for higher mutation rates $p_{mut}$ because they produce much more different parameters as mutations are the only way of introducing new parameters in the system. However, when looking at the relationship between the total number of parameters $|\mathcal{P}|$ and the mutation rates, as shown in figure \ref{fig:P_mut}, we can see that they relate  sub-linearly. This indicates the presence of competitive dynamics in the system, since without competition the number of parameters will necessarily increase linearly with the mutation rate. Interestingly, after a rapid growth, the number of different parameters tend to stay very stable through time and this for all the different values of $p_{mut}$ also indicating the presence of intrinsic regulation mechanisms. \\
When measuring evolutionary activity of the system, we also notice large differences when varying the mutation rates in the vanilla setting as shown in figure \ref{fig:ea_mut}. Importantly, results show a rapid decline of evolutionary activity measures for greater mutation rates where the relation is best fitted by a power law with slope $\gamma=-0.5$ ($R^2=0.75$) for non-neutral activity and $\gamma=-0.71$ ($R^2=0.71$) for count-based activity. These results are consistent with results from \citet{droop2012} where they observed declines of evolutionary activity for too high mutation rates. When expressing the evolutionary activities as a function of time for different mutation rates, all of the curves are better fitted by a linear model. The slopes of these fitted curves are higher for low mutation rates and  when plotting the slope of the curve over the mutation rate (figure \ref{fig:ea_mut}) we can observe a decay following a power function.\\

\begin{figure}[tp] % Example Figure
    \centering
    \includegraphics[width=0.9\textwidth]{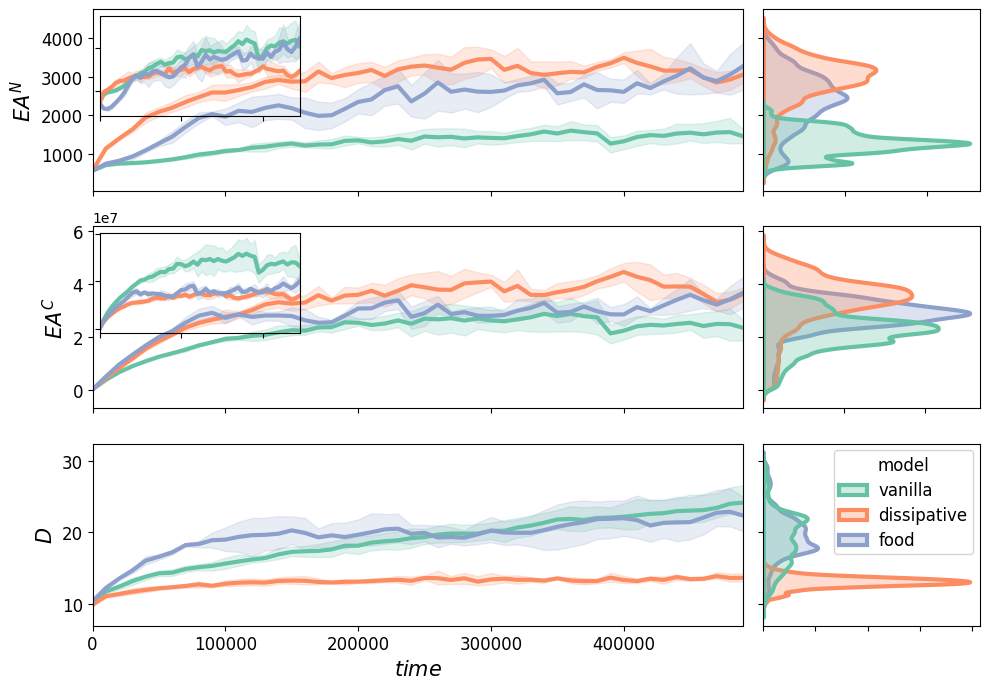}
    \caption{Comparison of non-neutral (top), count-based (middle) evolutionary activities and diversity (bottom) for the vanilla, dissipative and food models. (left) Metrics are plotted as a function of time for each model. For the evolutionary activity, inner plots shows the corrected measures (divided by total mass in the environment). (right) Distributions of respective metrics for each model (estimated through kernel density estimation).}
    \label{fig:EA_mdls}
\end{figure}

When comparing the vanilla, dissipative and food models respective evolutionary activities we observe that dissipative and food models display significantly higher evolutionary activities for both count-based and non-neutral measures ($p<10^{-5}$ for both, Mann-Whitney test). The dissipative model also shows higher evolutionary activity than the food model ($p<10^{-5}$ for both, Mann-Whitney test). However, one should note that in the case of the food and dissipative models, total mass in the environment is not ensured to stay constant. This creates a bias in the evolutionary activity measures which takes mass into account. When removing the total mass effect by dividing the evolutionary activity by the total mass in the environment, we obtain the opposite relationships. We especially observe much lower measures in the dissipative case in comparison to the other two models. While the vanilla model shows higher score than the food model when corrected activity is averaged over all time steps, we can see similar values in the end of simulations indicating a slower rise for the food model. Diversity measures (see figure \ref{fig:EA_mdls}, bottom) show striking differences where the dissipative model produces much less diversity than the other two. This might seem counter-intuitive as of the three it is the only one inputing new parameters. However, since the parameter map is never regularized, intrinsically evolving parameters can take any value which might be way out of the distribution from which we sample new parameters ($\mathcal{N}(0,1)$). The food model shows a quicker rise in diversity in comparison to the vanilla model but also stabilizes quicker while the vanilla model displays a steady constant linear growth of diversity.

\section{Discussion}\label{sec:disc}

Due to the mass conservative nature of Flow Lenia, most of patterns do not grow indefinitely into spatially global patterns (i.e patterns that diffuse on the entire grid, also called Turing-like patterns), therefore SLPs are much more common and easier to find. This is an important difference from the previous versions of Lenia, where one needs to search or evolve for patterns that are both non-vanishing and non-exploding, and to constantly monitor their existential status. Here, the mass conservation constraint acts as a regularizer on the kinds of patterns that can emerge. \\
Even though patterns generated by Flow Lenia are often static or slowly moving, we have been able to find creatures with complex dynamics from random search only which would be a difficult task in Lenia as most of the search space corresponds to either exploding or vanishing patterns. Furthermore, we have shown that the update rule parameters can be optimized with simple evolutionary strategies to generate patterns with specific properties and behaviors such as locomotion, chemotaxis and navigation through obstacles. Doing so in Lenia is a difficult task since the spatial localization of emergent patterns is not guaranteed necessitating more complex algorithms accounting for such a property.\\
Finally we showed that the Flow Lenia system allowed for the integration of the update rule parameters within the CA dynamics allowing for the coexistence of multiple update rules, and thus different creature or species, within the same simulation. The quantitative and qualitative analysis of trajectories of parameters through time, as well as the application of evolutionary activity metrics \citep{droop2012} and diversity metrics allowed us to shed light on the intrinsic evolution taking place in larger spatio-temporal scale simulations of this system. Intrinsic evolution, i.e evolution without externally defined stationary fitness function, is a particularly important feature of life as it supports its open-endedness through mechanisms such as niche construction. \\
We argue that such multispecies simulations represent an important step towards the design of emergent microcosms \citep{arbesman2022} in which could emerge intrinsic, maybe open-ended, evolutionary processes through inter-species interactions. Whereas environment design is poorly addressed and quite challenging in cellular automata systems, we believe that it is crucial to study the emergence of agency and cognition in those systems as argued in \citet{godfrey-smith2002} and shown in \citet{hamon2024}. By proposing different environment designs inspired by theories about origins of life and evolution, we have been able to study the evolutionary influences of such variations. By enabling the design of complex environmental features like inter-species interactions, walls, food or temperature, Flow-Lenia could represent a particularly interesting system to study theories on the origins of life or ecological theories of the evolution of complexity and cognition \citep{nisioti2021a,moulin-frier2022}.\\
Lot of exciting roads remain to be taken in order to fully capture the value of complex self-organized systems such as Flow-Lenia and explore their potential as models for studying  theories about life, cognition and evolution. While we showed that evolutionary activity metrics are applicable to the Flow-Lenia system with minimal modification, we have shown that EA measures are able to characterize the effect of different experimental conditions (vanilla, dissipative and food) on the resulting evolutionary dynamics. In evolutionary biology and theories on the origins of life, the presence of a dissipative mechanism as well of limited shared resources are considered as key drivers of open-ended evolution in the natural world. However, our results show an inverse tendency compared to these predictions, where the EA of the dissipative and food conditions are lower than in the vanilla systems. This is an interesting illustration of the interest of quantitive models such as Flow-Lenia, encouraging the formal definition of such mechanisms and measures. Our results show that common predictions on the origins of life actually strongly depends on their specific instantiation, since in our current model we actually observe the inverse tendency. However, we think that more improvements have to be made here. In particular, our working definition of species, i.e a specific point in the parameter space, might not fully capture what species are in our system. Looking at the distribution of parameters revealed the ubiquitous presence of clusters coherently moving in the high dimensional space of parameters with occasional divisions (i.e branchings). We believe that the correct definition of a species in Flow-Lenia might lie in these coherent clusters which might, through the scope of evolutionary activity metrics, shed new light on Flow-Lenia evolutionary dynamics. Another important contribution could also be to take into account the phenotypic outcomes of the parameters in the definition of species. Beside the notion of species, the definition of individuals could be reframed, or refined, as we often observe in simulations stable structures one would identify as an individual creature in the system but which are composed of elements defined by different set of parameters. While evolutionary theories have for a long time thought about genes as the fundamental unit of selection, recent theories propose that the individual, or agent, should be seen as the fundamental unit of evolution \citep{levin2023a}. Further study using Flow-Lenia might benefit investigating new methods for defining these individuals, for example using information theoretic measures of individuality \citep{krakauer2020}. Also, while measures of evolutionary activity represent a great tool for getting insights into the complex emerging dynamics of Flow-Lenia, other measures could be adapted and used in this setting. For instance, \citeauthor{patarroyo2023} proposed a framework based on assembly theory \citep{sharma2023} aimed at quantifying the open-endedness of discrete CA \citep{patarroyo2023}.\\
In conclusion, we believe that Flow-Lenia represents an important step towards the realization of open-ended evolutionary dynamics \emph{in silico}. By enabling great diversities of creatures to emerge, interact and evolve in complex large-scale environments, Flow-Lenia could lead to the emergence of complex cognition in the most as-it-could-be sense of the term. These emerging dynamics might shed new lights on studies about origins of life and cognition.

\printbibliography

\end{document}